# Orbital-enhanced Warping Effect in $P_xP_y$-derived Rashba Spin Splitting of Monatomic Bismuth Surface Alloy

*Guan-Yu Chen, Angus Huang, Yen-Hui Lin, Chia-Ju Chen, Deng-Sung Lin, Po-Yao Chang, Horng-Tay Jeng\*, Gustav Bihlmayer and Pin-Jui Hsu\**

G. Y. Chen, Dr. A. Huang, Y. H. Lin, C. J. Chen, Prof. D. S. Lin, Prof. Po-Yao Chang, Prof. H. T. Jeng, Prof. P. J. Hsu
Department of Physics, National Tsing Hua University, 30013 Hsinchu, Taiwan
E-mail: jeng@phys.nthu.edu.tw; pinjuihsu@phys.nthu.edu.tw

Prof. D. S. Lin, Prof. H. T. Jeng, Prof. P. J. Hsu
Center for Quantum Technology, National Tsing Hua University, Hsinchu 30013, Taiwan

Prof. H. T. Jeng
Physics Division, National Center for Theoretical Sciences, Hsinchu 30013, Taiwan

Prof. H. T. Jeng
Institute of Physics, Academia Sinica, Taipei 11529, Taiwan

Dr. Gustav Bihlmayer
Peter Grünberg Institute and Institute for Advanced Simulation, Forschungszentrum Jülich and JARA, 52425 Jülich, Germany

Abstract:

Spin-split Rashba bands have been exploited to efficiently control the spin degree of freedom of moving electrons, which possesses a great potential in frontier applications of designing spintronic devices and processing spin-based information. Given that intrinsic breaking of inversion symmetry and sizeable spin-orbit interaction, two-dimensional (2D) surface alloys formed by heavy metal elements exhibit pronounced Rashba-type spin splitting on surface states. Here, monatomic $BiCu_2$ surface alloy has been investigated by low-temperature scanning tunneling spectroscopy (LT-STS) combined with density functional theory (DFT). From differential conductance dI/dU spectra together with calculated electronic band structures, occupied and unoccupied Rashba spin-split surface bands originated from different hybridization of atomic orbitals have been resolved. By means of quasi-particle interference (QPI) measurements, energy-dependent standing wave patterns developed by two distinct scattering vectors along both $\overline{\Gamma M}$ and $\overline{\Gamma K}$ directions have been clearly observed in the dispersion of $p_xp_y(m_j = 1/2)$ band. According to spin-dependent constant energy contours (CECs) from DFT, essential out-of-plane spin component $S_z$ as a result of in-plane potential gradient on $BiCu_2$ surface reconstruction has been uncovered, which leads to extra opening channels for scattering transitions between time-reversal partners in considerably warped $p_xp_y(m_j = 1/2)$ Rashba spin-split band.

## 1. Introduction

Due to strong spin-orbit coupling, Dresselhaus pointed out that the spin degeneracy of electronic band structure can be lifted in bulk crystals lacking of inversion symmetry.[1] Same analogy can be applied to the surface or interface where the inversion asymmetry naturally exists, Rashba and Bychkov further conducted the pioneer work on the spin-orbit driven spin splitting of electronic band structure in two-dimensional (2D) systems.[2] Such spin splitting is intrinsic and particularly significant in elements with large atomic number, which also gives rise to an important feature of spin-momentum locking leading to an absence of backscattering in transport electrons.[3,4] Since an external magnetic field is no longer needed for this reduction of energy dissipating channel, Rashba-type spin splitting is then of great interest to technology application in this regard, one famous proposal belongs to the concept of Datta-Das spin transistor, which has promoted many efforts to design and pursuit Rashba effect on low-dimensional materials.[5]

By means of evaporating heavy atoms onto the noble metal surfaces, the 2D binary Rashba alloys can be successfully prepared, for examples, Pb, Bi and Sb grown on Ag(111) and Cu(111).[6-14] With an amount of 1/3 monolayer (ML) Bi on Ag(111), the gigantic Rashba spin-splitting characterized by the Rashba parameter $\alpha_R$ of 3.05 has been reported on BiAg$_2$ alloy with $\sqrt{3} \times \sqrt{3}R30^0$ superstructure.[7,15] Due to considerable in-plane potential gradient from each Bi atom surrounded by six Ag atoms in $\sqrt{3} \times \sqrt{3}R30^0$ lattice arrangement, BiAg$_2$ surface alloy also exhibits substantial out-of-plane spin polarization in addition to the spin-momentum locking of in-plane components. Furthermore, an important figure of merit for practical application, this 2D BiAg$_2$ alloy can be conveniently transferred to grow on semiconductor Si(111) substrate, i.e., the most common and compatible material used in industrial manufacturing, and a pronounced circular dichroic effect emerges as a result of the coupling between Rashba spin-split surface state and quantum-well states of Ag films.[8]

These interesting phenomena and properties of one-layer-thick BiAg$_2$ binary alloy have mostly been investigated on the Rashba band with atomic orbital hybridization of Ag 5$s$ and Bi 6$p_z$, which typically can be accessed in energy region below Fermi level, i.e., occupied states of valence band, by employing the photoemission spectroscopy. Besides $sp_z$ hybridization, however, there are also Rashba bands hybridized from different atomic orbitals, e.g., the Bi 6$p_x p_y$($m_j$ = 3/2; 1/2) and 6$p_z$ locating at energy region of unoccupied states above Fermi level. In order to study their Rashba characteristics, either depositing of various adsorbates, e.g., Na, Xe and Ar etc., to shift unoccupied Rashba bands downward to below Fermi energy or performing inverse photoemission experiment was required.[11] While a wealth of studies and understandings have been intensively focused on occupied $sp_z$ Rashba band, only few reports and relatively less are known for the unoccupied $p_x p_y$ and $p_z$ -derived Rashba-split surface states in various binary surface alloys.

With high spatial and energy resolutions, tunneling spectroscopy measurement offers an alternative approach to explore Rashba physics in both occupied and unoccupied energy ranges.[6,7] According to energy-dependent quasiparticle interference (QPI) experiment, the $sp_z$ and $p_x p_y$($m_j$ = 1/2)-derived Rashba bands in BiAg$_2$ alloy have been mapped out and the backscattering events above and below the Rashba energy $E_R$ are allowed under the assumption of the Bloch states.[16] From the dispersion of scattering vectors, the unoccupied $p_x p_y$($m_j$ = 1/2) band shows the spin-orbit entangled spin texture and hybridizes with occupied $sp_z$ band leading to either a gap opening from strong spin-orbit coupling or inter-band spin orbit coupling between even and odd-orbital components.[16-18] Analogous to studies on BiAg$_2$ alloy, same experimental technique can also be applied to access the two unoccupied Rashba-split bands of PbAg$_2$ surface alloy.[6] Two asymmetric peaks observed at unoccupied states in tunneling spectra clarify the larger inter-band separation, and quasiparticle interference mappings further confirm the absence of hybridization of these two unoccupied bands.[6]

In this work, we study the unoccupied Rahba-split surface states of BiCu$_2$ surface alloy on Cu(111) by using low-temperature scanning tunneling spectroscopy (LT-STS) combined with density functional theory (DFT). Differential conductance dI/dU spectra on BiCu$_2$ show an asymmetric peak at 0.23 eV originated from the singularity of LDOS at band edge of Rashba-split band formed by Cu 4$s$ and Bi 6$p_z$ orbital hybridizations. Furthermore, dI/dU peak at 1.6 eV together with a shoulder at 1.4 eV are developed from the band crossing of four-fold degenerate $p_xp_y$($m_j$ = 3/2) band right at $\bar{\Gamma}$ point and the band edge of $p_xp_y$($m_j$ = 1/2) band, respectively. From energy-dependent QPI mappings, we have clearly observed standing wave patterns, and two distinct scattering vectors along both $\overline{\Gamma M}$ and $\overline{\Gamma K}$ are able to be extracted at energy region covering the dispersion of $p_xp_y$($m_j$ = 1/2) band. With substantial in-plane potential gradient from protrusion of Bi atoms on BiCu$_2$ surface, first-principle calculation reveals significant out-of-plane spin component $S_z$ in spin-dependent CECs. Such warping deformation results in opening scattering channels for possible transitions between time-reversal partners from $p_xp_y$($m_j$ = 1/2)-derived Rashba spin-split band.

## 2. Results and Discussion

### 2.1. Fabrication and Structure of Monatomic Bismuth Surface Alloy

According to structure models obtained from previous X-ray diffraction and STM studies, the surface alloys of Bi grown on Cu(111) have Bi coverages of 1/3 ML and 1/2 ML for $\sqrt{3} \times \sqrt{3}R30^0$ and [2012] phases, respectively.[19,20] This in principle provides the way how we can calibrate the amount of Bi in order to prepare the sample with well-defined and extended $\sqrt{3} \times \sqrt{3}R30^0$ BiCu$_2$ alloy. The overview STM topography of about 0.40 ML Bi on Cu(111) has been shown in **Figure 1**(a) and the magnified image of Figure 1(b) from square area framed by black dashed lines in Figure 1(a) displays one antiphase domain boundary (blue arrow) with few Bi vacancies (red arrow) on BiCu$_2$ phase. These two types of defects, i.e., antiphase domain boundary and Bi vacancies, can be served as scattering centres enabling us to study the dispersion of Rashba-split bands from energy-dependent QPI measurements.[13] From the atomic resolution image in Figure 1(c), the surface reconstruction exhibits a $\sqrt{3} \times \sqrt{3}$ periodicity with a $30^0$ rotation to the high symmetry axis, e.g., $[1\bar{1}0]$, of Cu(111) underneath, which is in agreement to corresponding structure model of BiCu2 alloy shown in the inset.[19,20] Since the Bi coverage of this sample is a bit higher than 1/3 ML, few isolated patches of [2012] phase also coexist with the $\sqrt{3} \times \sqrt{3}R30^0$ surface alloy. For example, the atomic resolution image of Bi[2012] shown in Figure 1(d) has been resolved from blue squared frame in Figure 1(a), and structure model in the inset of Figure 1(d) can explain zigzag chain like feature in atomic resolution image by a $60^0$ rotation with respect to $[1\bar{1}0]$ direction of Cu(111) lattice with three-fold symmetry. We note that a monolayer of extended Bi[2012] can be fabricated by further increasing the Bi coverage (not shown here), at which Mathias *et al*. have revealed a large spin-splitting of unoccupied quantum-well states.[21]

**2.2. Electronic Structures of Rashba Spin-split Bands**

In order to determine the energy positions of different Rashba-split bands of Bi/Cu(111), the tunneling spectra have been acquired on both $\sqrt{3} \times \sqrt{3}R30^0$ BiCu$_2$ and Bi[2012] phases. **Figure 2**(a) displays the conductance dI/dU curve taken at modest bias range around Fermi level, e.g., from 0.7 to -0.4 V, on BiCu$_2$ (black line), an asymmetric peak at 0.23 V is originated from the singularity of local density of states (LDOS) at the band edge of *sp$_z$*-derived Rashba band.[13] From the asymmetry of this peak, we can also identify that this Rashba-split band has a downward parabolic dispersion. The Rashba splitting, Rashba energy and Rashba parameter of this band along the high symmetry direction $\overline{\Gamma M}$ ($\overline{\Gamma K}$) derived from our first-principles calculations are $k_0 = 0.0347(0.0359)$, $E_R = \hbar^2 k_0^2/2m^* = 0.0156(0.0169)$ and $\alpha_R = \hbar^2 k_0/2m^* = 0.899(0.941)$, respectively. Note that our DFT Rashba energy $E_R = 16$ meV is similar to ~ 20 meV obtained from photoemission studies, but our fitting of this dI/dU curve (not shown) gives $E_R \sim 65$ meV, which might be due to the broadening of 20 meV modulation voltage used in our lock-in measurements.[7,22] In contrast to BiCu$_2$, dI/dU curve measured at Bi[2012] (blue line) is rather featureless within this bias range.

To explore Rashba-split bands hybridized from $p_xp_y$ orbitals on BiCu$_2$, we have measured the dI/dU curve at larger bias range from 2.0 to -0.4 V and the results are shown in Figure 2(b). Interestingly, a pronounced peak at 1.6 V as indicated by red arrow accompanied by a shoulder at 1.4 V (blue arrow) have been observed on BiCu$_2$ (black line), but no characteristic features can be observed from dI/dU curve taken at Bi[2012] (blue line). Since the intensity of conductance peak at 1.6 V is unusually higher than the Rashba peak at 0.23 V (black arrow), we have double checked the normalized dI/dU/(I/U) from simultaneously measured I/U curve. The intensity of conductance peak at 1.6 V is indeed about 6 times higher than that of Rashba peak at 0.23 V (see Supplementary Figure S1 for more details), but less profound as compared to the result output directly from Lock-in amplifier.

Apart from tunneling spectroscopy measurements, we also performed first principle electronic structure calculations to reveal the origin of spectra peaks observed at different energy positions. The orbital-decomposed band structure of the surface BiCu$_2$ ions are shown in Figure 2(c). It can be seen that two sets of Rashba-type bands cross the $\bar{\Gamma}$ point at 0.2 eV and 1.4 eV, which correspond to the observed minor peak and the shoulder in our STS in Figure 2(a) and (b). In addition, there exists another band crossing near 1.8 eV, being in good agreement with the major peak observed in our STS Figure 2(b). For a direct comparison with the zone-center sensitive STS measurement, we further integrate the DOS around the $\bar{\Gamma}$ point with suitable broadening as shown in Figure 2(d). Our DFT calculations fully support the minor peak at about 0.2 eV, and the major peak at ~1.6 eV as well as the shoulder at ~1.4 eV observed in our STS measurement.

## 2.3. QPI Mappings of Hexagonally Warped Rashba-split Surface States

To uncover the dispersion of $p_xp_y$-derived Rashba-split bands and compare their relations to $sp_z$-derived Rashba band studied from previous work, we have carried out the QPI measurements covering both energy ranges on $BiCu_2$.[13] As shown in STM topography image of **Figure 3**(a), the Bi vacancies and one antiphase domain boundary have been served as scattering centres to develop standing electron wave patterns. According to the theoretical calculation by Mirhosseini *et al.*, the $p_xp_y(m_j = 1/2)$-derived Rashba band exhibits an unconventional spin topology, the overall feature can be inferred from Figure 3(b) with two scattering vectors $q_n(E)$.[23] Note that the energy-dependent scattering vector $q_n(E)$ is defined by $q_n(E) = k_f(E) - k_i(E)$ under the framework of elastic scattering, which links initial and final momentum eigenstates for mapping out the dispersion relation of surface bands.[13,16,17]

Although spin-conserved scattering, i.e., $k_f(E)$ and $k_i(E)$ with the same direction of spin-polarization, is typically considered in analysis of Rashba-split band, we are also aware of spin-flip scattering which needs to be taken into account for the case of band structure with a spin-polarization inversion.[6,13,16,17,24] However, we denote that either spin-conserved or spin-slip scattering only gives rise to a single scattering vector responsible for standing wave patterns from $p_xp_y(m_j = 1/2)$-derived Rashba-split surface state.

Another feature worth being mentioned is the hexagonal shape of CEC on $p_xp_y(m_j = 1/2)$-derived Rashba band and this signature has been experimentally observed at both $BiAg_2$ and $BiCu_2$ surface alloys, which can be associated with a coupling between $\sqrt{3} \times \sqrt{3}R30^0$ crystalline structure and symmetry of $p_xp_y$ atomic orbitals.[15,25] According to hexagonal warping effect on topological insulators (TIs), multiple pairs of stationary $k$ points on warped CEC can contribute to additional scattering vectors for the emergence of standing wave patterns as verified by several QPI results before.[26-29,32] As shown in the schematic drawing at the bottom of Figure 3(b), the hexagonally warped CEC on $p_xp_y(m_j = 1/2)$-derived Rashba band will open scattering channels along high symmetry directions of $\overline{\Gamma M}$ and $\overline{\Gamma K}$ (red and blue arrow lines), which facilitates the scattering events between time-reversal partners despite of backscattering remains protected by time-reversal symmetry at certain *k* points (green dashed arrow lines).[27,33,34]

From Figure 3(c) to (h), a series of dI/dU mappings clearly represent spatial modulations of LDOS, i.e., so called Friedel oscillations from scattering off antiphase domain boundary and Bi vacancies, at bias voltages ranged from 1.4 V to 0.4 V.[35] Within this energy region, it is worth mentioning that there are electron standing waves with different wavelengths from $p_xp_y(m_j = 1/2)$-derived Rashba-split band, and more than one scattering vectors are able to be extracted from corresponding Fast-Fourier-Transformation (FFT) analyses. Note that the standing waves become barely visible when bias voltages are above 1.4 V. According to Fourier-transformed dI/dU (FT-dI/dU) mapping at 1.4 V shown at the bottom of Fig. 3(c), six diffraction spots marked by red circles originate from $\sqrt{3} \times \sqrt{3}R30^0$ lattice of BiCu$_2$, which can be used to identify $\overline{\Gamma M}$ and $\overline{\Gamma K}$ directions in reciprocal space. In addition to that, real-space standing waves produce separate diffraction patterns in FFT image as indicated by red and blue arrows, and they evolve as a function of bias voltage, e.g., the diffraction patterns shrink as the bias voltage increases, which enables us to trace the energy-dependent scattering vectors $q_n(E)$.

### 2.4. Rashba Band Dispersions and Spin-dependent CECs

**Figure 4**(a) summarizes energy-dependent $q_n(E)$ along $\overline{\Gamma M}$ within bias voltage range from 1.5 V to -0.45 V (see $q_n(E)$ along $\overline{\Gamma K}$ in Supplementary). For the band dispersions of $D_1$ (black dots) and $D_2$ (green dots), they are formed from intra- and interband scattering processes of $sp_z$ and $p_xp_y(m_j = 1/2)$ Rashba-split bands.[13] Note that the empty black and greed dots are the data points we extracted from previous studies for a direct comparison.[13] The black and green lines are the parabolic fittings to obtain the effective masses $m^*_{\overline{\Gamma M}}$ of (-0.34±0.02)$m_e$ and (-0.39±0.01)$m_e$, respectively. They are slightly larger than previous QPI results due to a bit of deviations from some data points, but they are still within the error bars. We denote that these $m^*_{\overline{\Gamma M}}$ values remain consistent with those reported from photoemission work.[22] Besides $D_1$ and $D_2$, QPI results also reveal the band dispersions of $D_3$ (gray dots) and $D_5$ (blue dots) at occupied energy region. From the overlapping of some blue dots with extracted blue empty dots, we refer the $D_5$ dispersion to $p_x p_y (m_j = 1/2)$ Rashba band from intraband transitions, i.e., $q_2$ in previous work.[13] Note that there are not many data points visible from intraband transitions of $p_x p_y (m_j = 1/2)$ Rashba band, in particular at occupied states, we speculate that this is because the scattering vectors are quite large, such that they correspond to very short wavelengths of standing waves in real-space, which might be hard to detect and easily buried into background noises. As for $D_3$ dispersion, we associate it with the scattering between outer branch of $sp_z$ and $p_x p_y (m_j = 1/2)$ bands, since it starts appearing at about 0.1 eV well above the projected bulk band gap of Cu(111) substrate, whereas the strong hybridization between outer branch of $sp_z$ band and bulk states does not play significant role to forbid the emergence of this scattering vector $q_3$.[13,14,25]

Interestingly, two dispersions of $D_4$ and $D_5$ have been resolved for $p_x p_y (m_j = 1/2)$ Rashba-split band, and they are composed of scattering vectors $q_4$ and $q_5$ connecting to the scattering transitions between time-reversal partners as shown in Figure 4(b). Typically, the Rashba-split band consists of two concentric circles on CEC with an in-plane chiral spin structure perpendicularly locked to the momentum vector, leading to prohibited backscattering as a consequence of time-reversal symmetry. However, this scenario is not perfectly true if there is a warping distortion on CEC of Rashba-split band, where not only single pair of stationary points can be stabilized, but also new scattering channels can be opened, similar to hexagonal warping effect reported on TI materials.[27,29-31,33]

In order to verify the warping behaviour of the $p_x p_y (m_j = 1/2)$ Rashba-split band, we further calculate the corresponding CECs with $S_z$ component at some representative energies as shown in Figure 4(c) to (f), which is the most important characteristic to recognize warping deformation of this band. From the Figure 4(c), the CEC of $p_x p_y (m_j = 1/2)$ Rashba-split band at 0.4 eV represents two hexagons colored by red and blue for positive and negative $S_z$ components, respectively. The hexagonally warped CEC creates multiple stationary points allowing scattering events to occur, for example, two black arrows refer to possible scatterings along high symmetry directions of $\overline{\Gamma M}$ and $\overline{\Gamma K}$ in Figure 4(c). As energy increases from 0.4 to 1.4 eV, i.e., approaching to the $p_x p_y (m_j = 1/2)$ band edge, the Rashba-split CEC turns into ring-like shape with reduced $S_z$ component, suggesting the substantial $S_z$ component as a direct consequence originated from the contribution of nontrivial warping term. For a comparison, we have also calculated the $S_z$ components of Rashba bands in $BiAg_2$ and found that $BiCu_2$ can have a polarization value $P_z$ about 3 times larger than that of $BiAg_2$ in $p_x p_y (m_j = 1/2)$ Rashba-split band. (see Supplementary Figure S4 for details)

## 3. Conclusion

In conclusion, we have studied unoccupied Rashb-split surface bands on $BiCu_2$ binary alloy through tunneling spectroscopy experiment combined with DFT theory. According to dI/dU spectra and calculated band dispersions, we have identified an asymmetric peak at 0.23 eV from the band edge of $sp_z$ Rashba band, and a strong peak at 1.6 eV with a shoulder at 1.4 eV from the band crossing at $\overline{\Gamma}$ point of four-fold degenerate $p_x p_y (m_j = 3/2)$ band and the band edge of $p_x p_y (m_j = 1/2)$ band, respectively. The energy-dependent QPI mappings clearly show standing waves evolved from $p_x p_y (m_j = 1/2)$ band dispersion and two distinct scattering vectors, $q_4$ and $q_5$ are able to be extracted from FFT diffraction patterns along both $\overline{\Gamma M}$ and $\overline{\Gamma K}$ directions. Given that about 1.0 Å protrusion of Bi atoms on $BiCu_2$ surface, the in-plane surface potential gradient results in considerable out-of-plane spin component $S_z$ as observed in DFT CECs, e.g., up to 3 times larger than BiAg$_2$ in the dispersion of $p_xp_y(m_j = 1/2)$ hybridized band, due to warping effect of in-plane symmetry of atomic orbitals coupled to surface crystalline structure. Such $S_z$ component in the warping deformation of CEC opens extra channels for scattering transitions between time-reversal partners of $p_xp_y(m_j = 1/2)$-derived Rashba-split band. Although backscattering remains forbidden and protected by time-reversal symmetry, our results reveal that hexagonally warped CEC is allowed to facilitate additional scattering events in Rashba-split surface states, which provides an interesting aspect of engineering relevant spin-dependent transport on low-dimensional Rashba materials with strong spin-orbit interaction.

## 4. Experimental Section

*Experimental Methods*: The Bi/Cu(111) were prepared in an ultrahigh vacuum (UHV) chamber with the base pressure below p ≤ 2×10$^{-10}$ mbar. The clean Cu(111) surface was first prepared by cycles of Ar$^+$ ion sputtering with an ion energy of 500 eV at room temperature and subsequent annealing up to 800 K. The Bi source with purity of 99.999% (Goodfellow) was e-beam sublimated onto Cu(111) surface at elevated temperature of 400K at which the well-defined and extended $\sqrt{3} \times \sqrt{3}R30^0$ BiCu$_2$ binary alloy can be grown. After preparation, the sample was immediately transferred into a low-temperature scanning tunneling microscope (LT-STM) from Unisoku Co. Ltd. (operation temperature T ≈ 4.5 K). The topography images were obtained from the constant-current mode with the bias voltage U applied to the sample. For scanning tunneling spectroscopy (STS) measurements, a small bias voltage modulation was added to U (frequency $v$ = 3991 Hz), such that tunneling differential conductance dI/dU spectra as well as dI/dU maps can be acquired by detecting the first harmonic signal by means of a lock-in amplifier.

*Theoretical Calculations*: First-principles calculations are performed using the Vienna Ab initio Simulation Package (VASP) based on the density functional theory (DFT).[36-38] The projector-augmented-wave-type pseudopotential with the Ceperley-Alder and Perdew-Zunger (CA-PZ) type exchange-correlation functional are adopted in the local density approximation (LDA) calculations.[39-43] We consider the BiCu$_2$ monolayer on top of 9-layer Cu(111) $\sqrt{3} \times \sqrt{3}$ substrate to simulate our experimental system. The ion positions of the top 3 layers are optimized until the residue force is smaller than 0.02 eVÅ$^{-1}$. After the geometrical optimization, the buckling height of Bi ion of the BiCu$_2$ alloy is fixed at 1.0 Å as deduced from our experimental spectroscopy result. Spin-orbit coupling is included in the self-consistent calculations with the energy cutoff of 400 eV over the 12 × 12 × 1 **k**-mesh. The DOS are integrated over the zone-center near the $\bar{\Gamma}$ point with an additional broadening of 0.01 and 0.1 eV. The two-dimension energy contours of the Rashba bands around the $\bar{\Gamma}$ point are calculated over the 40 × 40 **k**-mesh and then interpolated over the 360 × 360 **k**-mesh.

**Supporting Information**

Normalized tunneling conductance curve on BiCu$_2$, dispersion of $q_n(E)$ along $\overline{\Gamma K}$ direction, energy and spin dependent CECs with $S_x$ and $S_y$ components, out-of-plane spin component $S_z$ of BiAg$_2$ and BiCu$_2$ surface alloys and FFT-QPI movie can be found in the Supporting Information, which is available from Wiley Online Library or from the author.

**Acknowledgements**

G.Y.C. and A.H. contributed equally to this work. D.S.L. and P.J.H. acknowledge support from the competitive research funding from National Tsing Hua University, Ministry of Science and Technology of Taiwan under Grants No. MOST-108-2636-M-007-002 and MOST-107-2112-M-007-001-MY3, and center for quantum technology from the featured areas research center program within the framework of the higher education sprout project by the Ministry of Education (MOE) in Taiwan. H.T.J. also acknowledges support from NCHC, CINC-NTU, iMATE-AS, and CQT-NTHU-MOE, Taiwan.

Received: ((will be filled in by the editorial staff))
Revised: ((will be filled in by the editorial staff))
Published online: ((will be filled in by the editorial staff))

**Keywords**

hexagonal warping effect, $p_x p_y$-derived Rashba spin splitting, monatomic bismuth surface alloy, quantum interference mapping, spin-dependent constant energy contours

References

[1] G. Dresselhaus, *Phys. Rev.* **1955**, *100*, 580.

[2] Y. A. Bychkov, E. I. Rashba, *JETP Lett*. **1984**, *39*, 78-81.

[3] G. Bihlmayer, O. Rader, R. Winkler, *New J. Phys*. **2015**, *17*, 050202.

[4] P. Gambardella, I. M. Miron, *Phil. Trans. R. Soc. A* **2011**, *369*, 3175-3197.

[5] S. Datta, B. Das, *Appl. Phys. Lett.* **1990**, *56*, 665-667.

[6] L. El-Kareh, G. Bihlmayer, A. Buchter, H. Bentmann, S. Blügel, F. Reinert, M. Bode, *New J. Phys*. **2014**, *16*, 045017.

[7] C. R. Ast, G. Wittich, P. Wahl, R. Vogelgesang, D. Pacilé, M. C. Falub, L. Moreschini, M. Papagno, M. Grioni, K. Kern, *Phys. Rev. B* **2007**, *75*, 201401(R).

[8] G. Bian, L. Zhang, Y. Liu, T. Miller, C. T. Chiang, *Phys. Rev. Lett*. **2012**, *108*, 186403.

[9] L. Moreschini, A. Bendounan, I. Gierz, C. R. Ast, H. Mirhosseini, H. Höchst, K. Kern, J. Henk, A. Ernst, S. Ostanin, F. Reinert, M. Grioni, *Phys. Rev. B* **2009**, *79*, 075424

[10] S. N. P. Wissing, K. T. Ritter, P. Krüger, A. B. Schmidt, M. Donath, *Phys. Rev. B* **2015**, *91*, 201403(R).

[11] H. Bentmann, F. Reinert, *New J. Phys*. **2013**, *15*, 115011.

[12] H. Bentmann, T. Kuzumaki, G. Bihlmayer, S. Blügel, E. V. Chulkov, F. Reinert, K. Sakamoto, *Phys. Rev. B* **2011**, *84*, 115426.

[13] M. Steinbrecher, H. Harutyunyan, C. R. Ast, D. Wegner, *Phys. Rev. B* **2013**, *87*, 245436.

[14] L. Moreschini, A. Bendounan, H. Bentmann, M. Assig, K. Kern, F. Reinert, J. Henk, C. R. Ast, M. Grioni, *Phys. Rev. B* **2009**, *80*, 035438.

[15] C. R. Ast, J. Henk, A. Ernst, L. Moreschini, M. C. Falub, D. Pacilé, P. Bruno, K. Kern, M. Grioni, *Phys. Rev. Lett*. **2007**, *98*, 186807.

[16] L. El-Kareh, P. Sessi, T. Bathon, M. Bode, *Phys. Rev. Lett*. **2013**, *110*, 176803.

[17] S. Schirone, E. E. Krasovskii, G. Bihlmayer, R. Piquerel, P. Gambardella, A. Mugarza, *Phys. Rev. Lett*. **2015**, *114*, 166801.

[18] R. Noguchi, K. Kuroda, K. Yaji, K. Kobayashi, M. Sakano, A. Harasawa, T. Kondo, F. Komori, S. Shin, *Phys. Rev. B* **2017**, *95*, 041111(R).

[19] D. Kaminski, P. Poodt, E. Aret, N. Radenovic, E. Vlieg, *Surf. Sci.* **2005**, *575*, 233-246.

[20] Y. Girard, C. Chacon, G. de Abreu, J. Lagoute, V. Repain, S. Rousset, *Surf. Sci.* **2013**, *617*, 118-123

[21] S. Mathias, A. Ruffing, F. Deicke, M. Wiesenmayer, I. Sakar, G. Bihlmayer, E. V. Chulkov, Y. Koroteev, P. M. Echenique, M. Bauer, M. Aeschlimann, *Phys. Rev. Lett.* **2010**, *104*, 066802.

[22] A. A. Ünal, A. Winkelmann, C. Tusche, F. Bisio, M. Ellguth, C. T. Chiang, J. Henk, J. Kirschner, *Phys. Rev. B* **2012**, *86*, 125447.

[23] H. Mirhosseini, J. Henk, A. Ernst, S. Ostanin, C. T. Chiang, P. Yu, A. Winkelmann, J. Kirschner, *Phys. Rev. B* **2009**, *79*, 245428.

[24] H. Hirayama, Y. Aoki, C. Kato, *Phys. Rev. Lett*. **2011**, *107*, 027204

[25] H. Bentmann, F. Forster, G. Bihlmayer, E. V. Chulkov, L. Moreschini, M. Grioni, F. Reinert, *EPL* **2009**, *87*, 37003.

[26] Y. L. Chen, J. G. Analytis, J. H. Chu, Z. K. Liu, S. K. Mo, X. L. Qi, H. J. Zhang, D. H. Lu, X. Dai, Z. Fang, S. C. Zhang, I. R. Fisher, Z. Hussain, Z. X. Shen, *Science* **2009**, 325, 178.

[27] L. Fu, *Phys. Rev. Lett.* **2009**, *103*, 266801.

[28] K. Kuroda, M. Arita, K. Miyamoto, M. Ye, J. Jiang, A. Kimura, E. E. Krasovskii, E. Chulkov, H. Iwasawa, T. Okuda, K. Shimada, Y. Ueda, H. Namatame, M. Taniguchi, *Phys. Rev. Lett.* **2010**, *105*, 076802.

[29] T. Zhang, P. Cheng, X. Chen, J. F. Jia, X. C. Ma, K. He, L. L. Wang, H. J. Zhang, X. Dai, Z. Fang, X. C. Xie, Q. K. Xue, *Phys. Rev. Lett.* **2009**, *103*, 266803.

[30] P. Roushan, J. Seo, C. V. Parker, Y. S. Hor, D. Hsieh, D. Qian, A. Richardella, M. Z.

Hasan, R. J. Cava, A. Yazdani, *Nature* **2009**, *103*, 266803.

[31] Z. Alpichshev, J. G. Analytis, J. H. Chu, I. R. Fisher, Y. L. Chen, Z. X. Shen, A. Fang, A. Kapitulnik, *Phys. Rev. Lett.* **2010**, *104*, 016401.

[32] P. Sessi, M. M. Otrokov, T. Bathon, M. G. Vergniory, S. S. Tsirkin, K. A. Kokh, O. E. Tereshchenko, E. V. Chulkov, M. Bode, *Phys. Rev. B* **2013**, *88*, 161407(R).

[33] W. C. Lee, C. J. Wu, D. P. Arovas, S. C. Zhang, *Phys. Rev. B* **2009**, *80*, 245439.

[34] P. Rakyta, A. Pályi, J. Cserti, *Phys. Rev. B* **2012**, *86*, 085456.

[35] J. Friedel, *Nuovo Cimento. Suppl.* **1958**, *7*, 287-311.

[36] G. Kresse, J. Hafner, *Phys. Rev. B* **1993**, *48*, 13115.

[37] G. Kresse, J. E. Furthmüller, *Computational Materials Science* **1996**, *6*, 15.

[38] G. Kresse, J. E. Furthmüller, *Phys. Rev. B* **1996**, *54*, 11169.

[39] P. E. Blöchl, *Phys. Rev. B* **1994**, *50*, 17953.

[40] G. Kresse, D. Joubert, *Phys. Rev. B* **1999**, *59*, 1758.

[41] D. M. Ceperley, B. J. Alder, *Phys. Rev. Lett*. **1980**, *45*, 566.

[42] J. P. Perdew, A. Zunger, *Phys. Rev. B* **1981**, *23*, 5048-5079.

[43] W. Kohn, L. J. Sham, *Phys. Rev.* **1965**, *140*, A1133.

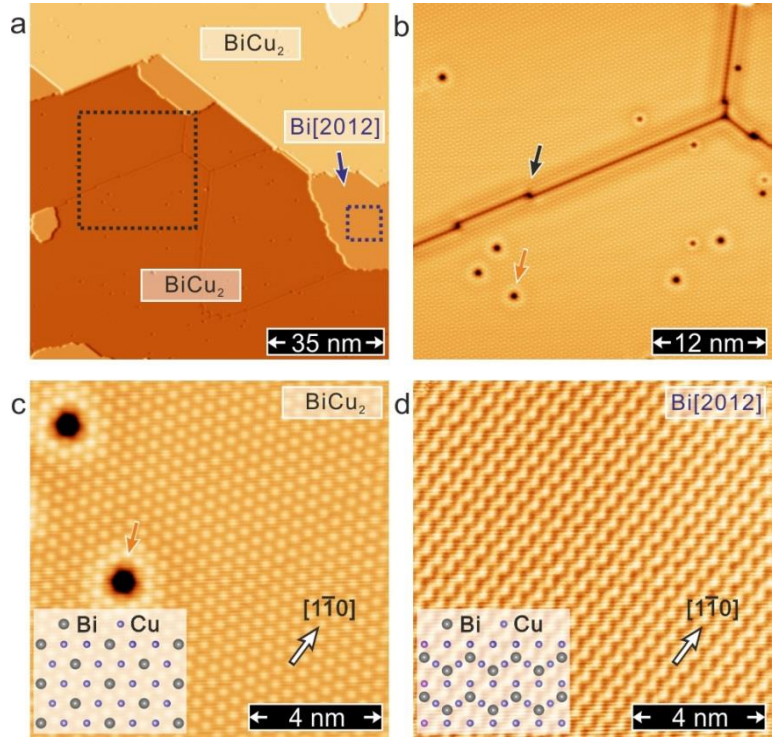

**Figure 1.** a) Overview of STM constant-current topography of about 0.40 ML Bi grown on Cu(111). (scan parameters: U = +1.0 V, I = 0.4 nA) b) Zoom-in image measured from the BiCu$_2$ area marked by black dashed square in a). The antiphase domain boundary and Bi vacancies are indicated by blue and red arrows, respectively. (scan parameters: U = +1.0 V, I = 1.0 nA) c) Atomic resolution image of BiCu$_2$ alloy shows $\sqrt{3} \times \sqrt{3}R30^0$ reconstruction and corresponding structure model has been shown in the inset. Red arrow marks the same single Bi vacancy from b). (scan parameters: U = +1.0 V, I = 1.0 nA) d) Atomic resolution image of Bi[2012] unit cell resolved from blue dashed square in a). The zigzag chain like feature can be inferred from a $60^0$ rotation with respect to $[1\bar{1}0]$ direction of structure model in the inset. (scan parameters: U = +1.0 V, I = 1.0 nA)

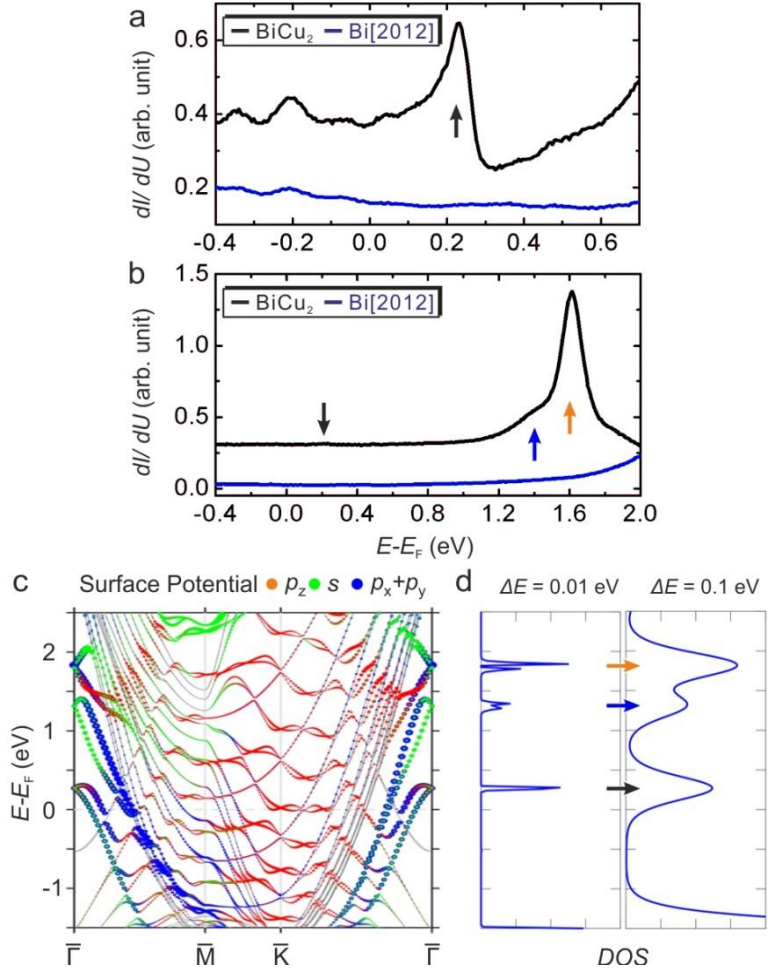

**Figure 2.** From a), an asymmetric peak at 0.23 V has been resolved from dI/dU curve on BiCu$_2$ (black line). In contrast to BiCu$_2$, the dI/dU curve on Bi[2012] is rather featureless (blue line). (stabilization parameters: U = +0.7 V, I = 1.0 nA). From b), a pronounced peak at 1.6 V as indicated by red arrow together with a shoulder at 1.4 V (blue arrow) have been observed on BiCu$_2$, but no characteristic features can be observed from dI/dU curve taken at Bi[2012]. (stabilization parameters: U = +2.0 V, I = 1.0 nA). c) Orbital-decomposed band structure of the surface BiCu$_2$ ions. The red, green, and blue spheres indicate respectively the $p_z$, $s$, and $p_x + p_y$ orbital contributions from the surface ions. The crossing points of the Rashba bands at $\bar{\Gamma}$ correspond to the peak positions in d). d) The DOS integrated over the zone center region near the $\bar{\Gamma}$ point for comparison with our zone-center sensitive STS results. The left (right) panel shows DOS with $\Delta E = 0.01$ eV ($\Delta E = 0.1$ eV) broadening.

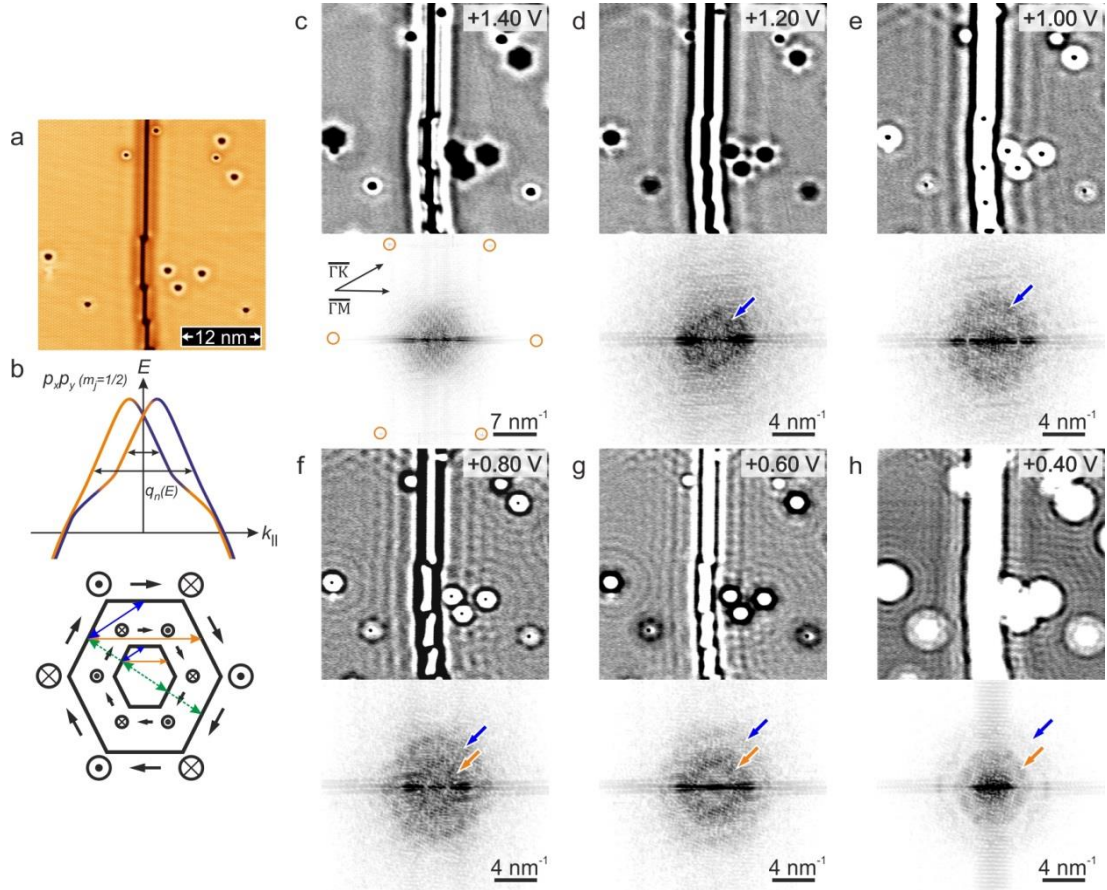

**Figure 3.** a) STM constant-current topography of BiCu$_2$ with an antiphase domain boundary and Bi vacancies used for scattering centres. (scan parameters: U = +1.0 V, I = 1.0 nA) b) Top: $p_xp_y(m_j = 1/2)$-derived Rashba-split band with unconventional spin topology, black arrow line refers to possible transitions between time-reversal partners. Bottom: schematic drawing of hexagonal CECs with opening scattering channels. Red and blue arrow lines are allowed scattering vectors along $\overline{\Gamma M}$ and $\overline{\Gamma K}$, respectively. Green dashed arrow line represents the forbidden backscattering processes. c)-h) dI/dU mappings at energy region covers $p_xp_y(m_j = 1/2)$ band. From the corresponding FFT images, we can identify high symmetry directions of $\overline{\Gamma M}$ and $\overline{\Gamma K}$ of BiCu$_2$ surface alloy (red circles in bottom of c)), and evolution of two scattering vectors within this energy region (red and blue arrows). (scan parameters: I = 1.0 nA for all dI/dU mappings)

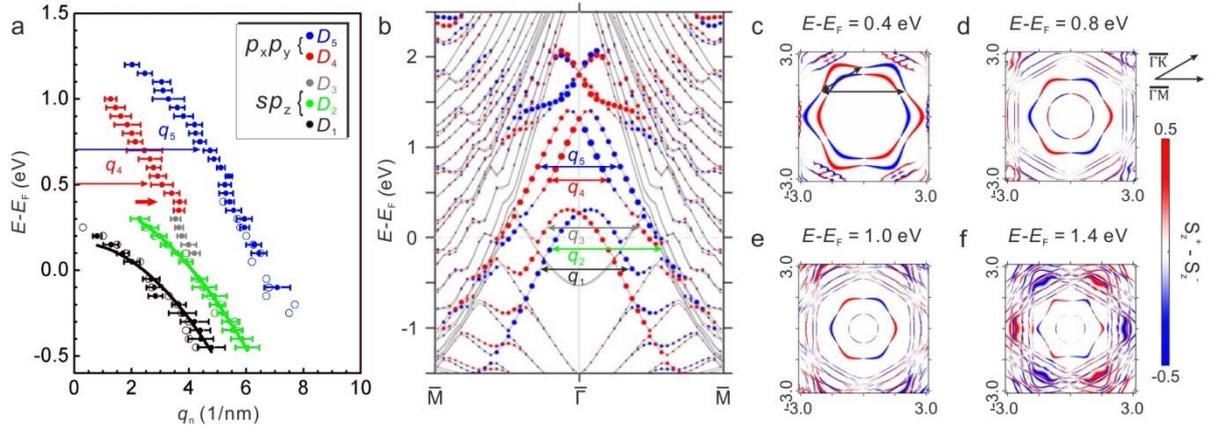

**Figure 4.** a) Dispersion of scattering vectors $q_n(E)$ along $\overline{\Gamma M}$ extracted from a series of dI/dU mappings from 1.5 V to -0.45 V. b) Spin-decomposed band structure of the surface $BiCu_2$ ions. The red (blue) circle indicates the in-plane spin up (down) component from the surface ions. Five possible $q$ vectors are shown by arrows in different colors. c)-f) Spin-resolved CECs at different energies given from DFT calculations. The red (blue) color indicates the out-of-plane spin up (down) component and hexagonal CECs reveals the warping effect.

**Hexagonally warped Rashba-split surface band develops considerable out-of-plane spin component** on constant energy Fermi contour, leading to scattering transitions allowed in between time-reversal partners. The first time observation of warping effect reported here on single atomic layer of bismuth surface alloy offers a superior prospect for efficiently designing spintronic devices and processing spin-based information on low-dimensional Rashba materials.

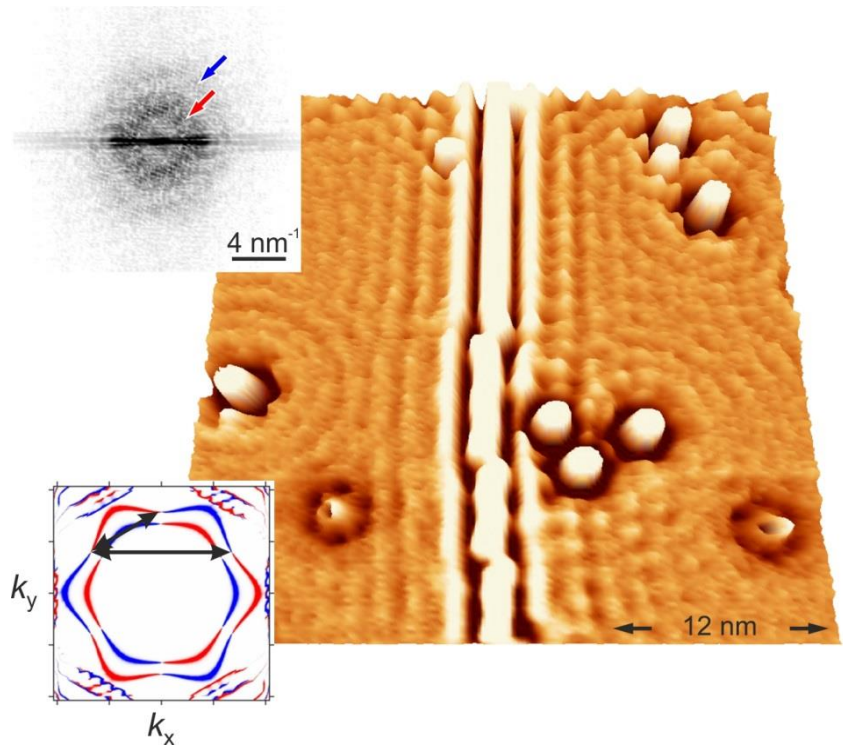